\documentclass[%
 reprint,
 amsmath,amssymb,
 aps,prd,
floatfix,
]{revtex4-2}

\usepackage{graphicx}
\usepackage{dcolumn}
\usepackage{bm}
\usepackage{xcolor}
\usepackage{comment}
\usepackage{rotating}
\usepackage{multirow}

\newcommand{\araa}{Annu. Rev. Astron. Astrophys.} 
\newcommand{\apjl}{Astrophys. J. Lett.} 
\newcommand{\aap}{Astron. Astrophys.} 
\newcommand{\mnras}{Mon. Not. R. Astron. Soc.} 
\newcommand{\physrep}{Phys. Rep.} 
\newcommand{\pasa}{Publ. Astron. Soc. Aust.} 
\newcommand{\pasp}{Publ. Astron. Soc. Pac.} 
\newcommand{\ssr}{Space Sci. Rev.} 

\begin{document}

\title{The long-short GRB connection} 


\author{J. A. Rueda$^{1,2,3,4,5}$}
\author{L. Becerra$^{6,1}$}
\author{C. L. Bianco$^{1,2,5}$}
\author{M. Della Valle$^{7,1}$}
\author{C. L. Fryer$^{8,9,10,11,12}$}
\author{C. Guidorzi$^{13,14}$}
\author{R. Ruffini$^{1,2,15,16}$}
\email{jorge.rueda@icra.it}

\affiliation{$^{1}$ICRANet, Piazza della Repubblica 10, I-65122 Pescara, Italy}

\affiliation{$^{2}$ICRA, Dipartimento di Fisica, Sapienza Università di Roma, Piazzale Aldo Moro 5, I-00185 Roma, Italy}

\affiliation{$^{3}$ICRANet-Ferrara, Dipartimento di Fisica e Scienze della Terra, Universit\`a degli Studi di Ferrara, Via Saragat 1, I-44122 Ferrara, Italy}

\affiliation{$^{4}$Dipartimento di Fisica e Scienze della Terra, Universit\`a degli Studi di Ferrara, Via Saragat 1, I--44122 Ferrara, Italy}

\affiliation{$^{5}$INAF, Istituto di Astrofisica e Planetologia Spaziali, Via Fosso del Cavaliere 100, 00133 Rome, Italy}

\affiliation{$^{6}$Centro Multidisciplinario de Física, Vicerrectoría de Investigación, Universidad Mayor, Santiago de Chile 8580745, Chile}

\affiliation{$^{7}$INAF - Osservatorio Astronomico di Capodimonte, Salita Moiariello 16, I-80131, Napoli, Italy}

\affiliation{$^{8}$Center for Theoretical Astrophysics, Los Alamos National Laboratory, Los Alamos, NM, 87545, USA}

\affiliation{$^{9}$Computer, Computational, and Statistical Sciences Division, Los Alamos National Laboratory, Los Alamos, NM, 87545, USA}

\affiliation{$^{10}$The University of Arizona, Tucson, AZ 85721, USA}

\affiliation{$^{11}$Department of Physics and Astronomy, The University of New Mexico, Albuquerque, NM 87131, USA}

\affiliation{$^{12}$The George Washington University, Washington, DC 20052, USA}

\affiliation{$^{13}$INFN - Sezione di Ferrara, Via Saragat 1, 44122 Ferrara, Italy}

\affiliation{$^{14}$INAF - Osservatorio di Astrofisica e Scienza dello Spazio di Bologna, Via Piero Gobetti 101, 40129 Bologna, Italy}

\affiliation{$^{15}$Universit\'e de Nice Sophia-Antipolis, Grand Ch\^ateau Parc Valrose, Nice, CEDEX 2, France}

\affiliation{$^{16}$INAF, Viale del Parco Mellini 84, 00136 Rome, Italy}

\begin{abstract}
Long and short gamma-ray bursts (GRBs) are thought to arise from different and unrelated astrophysical progenitors. The association of long GRBs with supernovae (SNe) and the difference in the distributions of galactocentric offsets of long and short GRBs within their host galaxies have often been considered strong evidence of their unrelated origins. Long GRBs have been thought to result from the collapse of single massive stars, while short GRBs come from mergers of compact object binaries. Our present study challenges this conventional view. We demonstrate that the observational properties, such as the association with SNe and the different galactic offsets, are naturally explained within the framework of the binary-driven hypernova model, suggesting an evolutionary connection between long and short GRBs.
\end{abstract}

\date{\today}


\maketitle

\section{Introduction}\label{sec:1}

The binary nature of short gamma-ray bursts (GRBs) was recognized and widely accepted since the first proposals based on mergers of binaries formed of two neutron stars (NS-NS) or an NS and a black hole (NS-BH; e.g. \cite{1986ApJ...308L..47G, 1986ApJ...308L..43P, 1989Natur.340..126E, 1991ApJ...379L..17N}). On the other hand, long GRBs have been mostly considered to arise from the core-collapse of a single massive star into a BH (or a magnetar), a \textit{collapsar} \cite{1993ApJ...405..273W}, surrounded by a massive accretion disk \cite{2002ARA26A..40..137M, 2004RvMP...76.1143P}. 

Therefore, the above theoretical models of long and short GRBs have treated them as two different and unrelated classes of astrophysical sources from different progenitors. This assumption has been further enhanced by the fact that only the long GRBs are associated with supernovae (SNe) and by the differences in the observed projected galactocentric offsets of short and long GRBs in the host galaxies. This work shows that such apparent differences are instead explained through an evolutionary connection between the long and the short GRBs that naturally arises when considering the role of binaries in the stellar evolution of massive stars.

Indeed, multiwavelength observations in the intervening years point to a key role of binaries in the evolution of massive stars and GRBs. The BeppoSAX satellite capabilities led to the discovery of the X-ray afterglow of GRBs \cite{1997Natur.387..783C}, and the accurate position, which allowed the optical follow-up by ground-based telescopes, led to two major results: determining the GRB cosmological nature \cite{1997Natur.387..878M} and observing long GRBs in temporal and spatial coincidence with type Ic SNe. The first GRB-SN association was GRB 980425-SN 1998bw \cite{1998Natur.395..670G}. The follow-up by the Neil Gehrels Swift Observatory \cite{2005SSRv..120..143B,2005SSRv..120..165B,2005SSRv..120...95R} of the optical afterglow has confirmed about twenty GRB-SN associations \cite{2006ARA&A..44..507W, 2011IJMPD..20.1745D, 2012grb..book..169H, 2017AdAst2017E...5C, 2023ApJ...955...93A}. The SNe Ic associated with the long GRBs show similar optical luminosity and peak time independent of the GRB energetics, which spans nearly seven orders of magnitude in the sample of GRB-SN (see Ref. \cite{2023ApJ...955...93A}, for details). Explaining the GRB-SN association is one of the most stringent constraints for GRB models. 

GRB-SN systems are related to massive star explosions \cite{2006Natur.441..463F,2008ApJ...689..358R,2008ApJ...687.1201K}, and most massive stars belong to binaries \cite{2007ApJ...670..747K,2012Sci...337..444S}. {The SN associated with long GRBs are of type Ic, and theoretical models and simulations show that} they are more plausibly explained via binary interactions to aid the hydrogen and helium layers of the pre-SN star to be ejected \cite{1988PhR...163...13N, 1994ApJ...437L.115I, 2007PASP..119.1211F, 2010ApJ...725..940Y, 2011MNRAS.415..773S, 2015PASA...32...15Y, 2015ApJ...809..131K}. {Further discussion on binary and single-star model progenitors of GRB-SNe can be found in \citet{2023ApJ...955...93A}.}

The above theoretical and observational considerations suggest that long GRBs {associated with SNe likely occur in binaries.} A possible crucial role of binaries in GRBs had been envisaged in  \citet{1999ApJ...526..152F}. The binary-driven hypernova (BdHN) model has proposed a binary progenitor for long GRBs to respond to {the above} exigences. In this model, the GRB-SN event arises from a binary comprising a carbon-oxygen (CO) star and an NS companion. The collapse of the {iron core of the} CO star leads to a newborn NS ($\nu$NS) and a type Ic SN. The explosion and expelled matter in the presence of the NS companion in a tight orbit triggers a series of physical processes that lead to the observed emission episodes (see, e.g., Refs. \cite{2012ApJ...758L...7R, 2015PhRvL.115w1102F, 2015ApJ...812..100B, 2016ApJ...833..107B, 2019ApJ...871...14B, 2022PhRvD.106h3004R, 2022PhRvD.106h3002B, 2023ApJ...955...93A}, and references therein). 
Most relevant is the hypercritical accretion of SN ejecta onto the $\nu$NS and NS companion \cite{2014ApJ...793L..36F}, {allowed by} the copious emission of MeV-neutrinos \cite{2016ApJ...833..107B, 2018ApJ...852..120B}. The accretion rate, highly dependent on the orbital period, leads to various BdHN types.

In the few-minute-orbital-period CO-NS binaries, the NS reaches the critical mass, collapsing into a rotating (Kerr) BH. These systems are called BdHN~I and are the most energetic long GRBs with an energy release $\gtrsim 10^{52}$ erg. Some examples are GRB 130427A \cite{2019ApJ...886...82R}, GRB 180720B \cite{2022ApJ...939...62R}, and GRB 190114C \cite{2021PhRvD.104f3043M, 2021A&A...649A..75M}. The accretion rate is lower in less compact binaries with periods from tens of minutes to hours, so the NS remains stable as a more massive, fast-rotating NS. These systems, called BdHNe~II, release energies $\sim 10^{50}$--$ 10^{52}$ erg. An example is GRB 190829A \cite{2022ApJ...936..190W}. Wide CO-NS binaries with periods of up to days, called BdHNe~III, release $\lesssim 10^{50}$ erg, such as GRB 171205A \cite{2023ApJ...945...95W}.

The above picture predicts that BdHN events (long GRBs) may lead to three possible fates of the CO-NS binary: an NS-BH (BdHNe~I) and NS-NS (BdHNe~II) or two runaway NSs (most BdHNe~III). The gravitational wave (GW) emission will lead the new compact-object binaries that remain bound to merge, producing short GRBs \cite{2015PhRvL.115w1102F, 2016ApJ...832..136R, 2018ApJ...859...30R, 2023Univ....9..332B}. We refer to this evolutionary process as the \textit{long-short GRB connection}. We have recently performed a suite of numerical simulations to determine the binary parameters that form NS-BH, NS-NS, and those that become unbound by BdHN events \cite{2024arXiv240115702B}. Here, we use those new results to assess the long-short GRB connection from the theoretical and observational viewpoint. In particular, we analyze information from the GRB density rates, the distribution as a function of redshift, the host galaxy types, and the projected offset position of long and short GRBs.

Section~\ref{sec:2} summarizes the observational constraints for the long-short GRB connection imposed by the observed GRB populations, density rates, the host galaxies, and the sources' position projected offsets. Section~\ref{sec:3} shows the main results of the three-dimensional numerical simulations of the BdHN scenario relevant to the analysis of this work. Specifically, we calculate the merger times and the difference of the position offsets between the long and short GRBs, predicted by the BdHN model simulations. In section~\ref{sec:5}, we discuss our results and draw the main conclusions.

\section{Observational constraints for the long-short GRB connection}\label{sec:2}

\subsection{GRB density rates}\label{sec:2.1}

A clue for the long-short GRB connection may arise from the GRB occurrence rates. {Here, we use the rates estimated in \citet{2016ApJ...832..136R}, following the method by \citet{2015ApJ...812...33S}. Suppose $\Delta N_i$ bursts are detected by various instruments in a logarithmic luminosity bin from $\log L$ to $\log L + \Delta \log L$. Thus, the total local density rate between observed minimum and maximum luminosities $L_{\rm min}$ and $L_{\rm max}$ can be estimated as
\begin{equation}\label{eq:rate}
    {\cal R} = \sum_i \sum_{L_{\rm min}}^{L_{\rm max}} \frac{4\pi}{\Omega_i T_i} \frac{1}{\ln 10}\frac{1}{g(L)}\frac{\Delta N_i}{\Delta \log L}\frac{\Delta L}{L},
\end{equation}
where $\Omega_i$ and $T_i$ are the instrument field of view and observing time, $g(L) = \int_0^{z_{\rm max}} (1+z)^{-1}dV(z)$, being $V(z)$ the comoving volume given in a flat $\Lambda$CDM cosmology by $dV(z)/dz = (c/H_0) 4\pi d_L^2/[(1+z^2)\sqrt{\Omega_M (1+z)^3)+\Omega_\Lambda}$, with $H_0$ the Hubble constant, $d_L$ the luminosity distance, $\Omega_M$ and $\Omega_\Lambda$ the cosmology matter and dark energy density parameters, and $z_{\rm max}$ is the maximum redshift at which a burst of luminosity $L$ can be detected. We refer the reader to Section $10$ in \cite{2016ApJ...832..136R} for further details.}

{Using a sample of $233$ long bursts with $E_{\rm iso} \gtrsim 10^{52}$ erg, peak energy $0.2\lesssim E_p\lesssim 2$ MeV, and measured redshifts $0.169\leq z\leq 9.3$, \citet{2016ApJ...832..136R} estimated the observed (isotropic) density rate of BdHN I, ${\cal R}_{\rm I}\approx 0.7$--$0.9$ Gpc$^{-3}$ yr$^{-1}$. As expected from the above definitions,} this rate agrees with {the estimated rate of the so-called} high-luminous {($L \gtrsim 10^{50}$ erg s$^{-1}$)} long GRBs, e.g., $0.6$--$1.9$~Gpc$^{-3}$~yr$^{-1}$ \cite{2010MNRAS.406.1944W} and $0.7$--$0.9$~Gpc$^{-3}$~yr$^{-1}$ \cite{2015ApJ...812...33S}.

{As discussed in \cite{2015PhRvL.115w1102F}, the} BdHN~I subclass can arise from a small subset of the ultra-stripped binaries. The rate of ultra-stripped binaries, ${\cal R}_{\rm USB}$, is expected to be $0.1\%$-–$1\%$ of the total SN Ic \cite{2015MNRAS.451.2123T}. The rate of SN Ic (not the total core-collapse SN) has been estimated to be ${\cal R}_{\rm SN Ic}\approx 2.6\times 10^{4}$ Gpc$^{-3}$ yr$^{-1}$ \cite[see, e.g.,][]{2007ApJ...657L..73G}. This estimate is compatible with more recent estimates, e.g., ${\cal R}_{\rm SN Ic}\sim 2.4\times 10^{4}$ Gpc$^{-3}$ yr$^{-1}$ \cite{2021MNRAS.500.5142F}. Therefore, the rate of ultra-stripped binaries may be ${\cal R}_{\rm USB}\sim 24$--$240$ Gpc$^{3}$ yr$^{-1}$, which implies that $\sim 0.4\%$-–$4\%$ of them may explain the BdHNe~I observed population.

\begin{table*}
\centering
\caption{Summary of some physical and observational properties of the GRB subclasses relevant for this work. The first three columns indicate the GRB subclass name and the corresponding pre-BdHN and post-BdHN binaries. In columns 4 and 5, we list the ranges of peak energy ($E_{\rm p,i}$) and isotropic energy released ($E_{\rm iso}$) (rest-frame $1$--$10^4$~keV). Columns 6 and 7 lists the maximum observed redshift and the local observed rate ${\cal R}$ obtained in \citet{2016ApJ...832..136R}.}\label{tab:rates}
\begin{tabular}{ccccccc}
     Subclass  & Pre-BdHN  & Post-BdHN & $E_{\rm p,i}$ &  $E_{\rm iso}$ &  $z_{\rm max}$ & ${\cal R}$ \\
     \hline
& & & (MeV) & (erg) & & (Gpc$^{-3}$yr$^{-1}$)\\  		
\hline
BdHN~I  & CO-NS  & NS-BH & $\sim0.2$--$2$ &  $\sim 10^{52}$--$10^{54}$ &  $9.3$ & $0.77^{+0.09}_{-0.08}$\\
BdHN~II+III & CO-NS    & NS-NS & $\lesssim 0.2$  &  $\sim 10^{48}$--$10^{52}$ &  $1.096$ & $100^{+45}_{-34}$\\
S-GRF & NS-NS & NS & $\lesssim2$ &  $\sim 10^{49}$--$10^{52}$ &  $2.609$ &  $3.6^{+1.4}_{-1.0}$\\
S-GRB  & NS-NS & BH & $\gtrsim2$ &  $\sim 10^{52}$--$10^{53}$ & $5.52$  & $\left(1.9^{+1.8}_{-1.1}\right)\times10^{-3}$\\
U-GRB & NS-BH & BH & $\gtrsim2$ &  $\gtrsim 10^{52}$ & $\lesssim z^{\text{I}}_{\text{max}}$ & $\lesssim {\cal R}_{\text{I}}$\\
\hline
\end{tabular}
\end{table*}

{Turning now to the BdHNe~II and III, the above method leads to the total density rate ${\cal R}_{\rm II+III} \approx 66$--$145$ Gpc$^{-3}$~yr$^{-1}$, which was estimated in \cite{2016ApJ...832..136R} with a sample of $10$ long bursts with $E_{\rm iso} \lesssim 10^{52}$ erg, $4\lesssim E_p\lesssim 200$ keV, and measured redshifts $0.0085 \leq z\leq 1.096$.
As expected from the above features, this rate agrees with independent estimates of the density rate of the so-called low-luminous ($L \lesssim 10^{48}$ erg s$^{-1}$)}  long GRBs, e.g., $148$--$677$~Gpc$^{-3}$~yr$^{-1}$ \cite{2007ApJ...662.1111L}, $155$--$1000$~Gpc$^{-3}$~yr$^{-1}$ \cite{2007ApJ...657L..73G}, $\sim 200$~Gpc$^{-3}$~yr$^{-1}$ \cite{2009MNRAS.392...91V}, and $99$--$262$~Gpc$^{-3}$~yr$^{-1}$ \cite{2015ApJ...812...33S}. Therefore, the BdHNe~II and III dominate the long GRB rate, i.e., ${\cal R}_{\text{long}}\equiv {\cal R}_{\text{I+II+III}} \approx {\cal R}_{\text{II+III}}$.

Let us now discuss the post-BdHN binaries formed by the BdHNe~I, II, and III. The (\textit{pre-BdHN}) CO-NS progenitors of BdHNe~I have orbital periods of a few minutes, so most of them remain bound after the SN explosion \cite{2015PhRvL.115w1102F, 2019ApJ...871...14B}. The bursts from {the NS-BH mergers formed after BdHNe I} are expected to have compact and potentially low-mass disks, leading to very short durations. Hence, they have been called ultra-short GRBs (U-GRBs). The above properties make U-GRBs hard to detect, and it is thought that no U-GRB has been observed \cite{2015PhRvL.115w1102F}. Thus, we can assume the rate of BdHN I as the upper limit to the U-GRBs from NS-BH mergers, i.e., ${\cal R}_{\text{U-GRB}}\lesssim {\cal R}_{\text{I}}$.

In BdHNe II and III, the SN can either disrupt the binary, leading to runaway NSs, or, if it remains bound, to an NS-NS binary. The mergers of the NS-NS binaries are expected to produce short GRBs. As for BdHN~I and II energy separatrix of $\sim 10^{52}$ erg related to the energy required to bring the NS companion to the critical mass for BH formation, in \citet{2016ApJ...832..136R, 2018ApJ...859...30R}, two subclasses of short bursts from NS-NS mergers have been distinguished. The mergers that overcome the NS critical mass, so those forming a BH, should release an energy $\gtrsim 10^{52}$ erg. These systems have been called authentic short GRBs (S-GRBs). The NS-NS mergers leading to a stable, massive NS have been called short gamma-ray flashes (S-GRFs) and release $\lesssim 10^{52}$ erg. It has been there estimated that ${\cal R}_{\text{S-GRF}}\approx 4$ Gpc$^{3}$ yr$^{-1}$ and ${\cal R}_{\text{S-GRB}}\approx 0.002$ Gpc$^{3}$ yr$^{-1}$. Hence, the S-GRFs dominate the rate of short bursts, i.e., ${\cal R}_{\text{short}}\equiv {\cal R}_{\text{S-GRF}}+{\cal R}_{\text{S-GRB}} + {\cal R}_{\text{U-GRB}} \approx {\cal R}_{\text{S-GRF}}$. This implies that NS-NS mergers dominate the observed local short GRB rate. The above estimates agree with independent assessments of the short GRB rate \cite[see Table 2 in][for a summary]{2018MNRAS.481.4355D}, and the current upper limit of AT 2017gfo kilonova-like events $< 900$ Gpc$^{3}$ yr$^{-1}$ \cite{2021ApJ...918...63A}. We refer to \citet{2022LRR....25....1M} for a recent review.

We summarize in Table~\ref{tab:rates} all the above information for the various BdHN and short GRB subclasses. The fact that ${\cal R}_{\text{long}}>{\cal R}_{\text{ short}}$ supports the expectation that the SN event disrupts a non-negligible fraction of binaries. Indeed, if we require the short-burst population to derive from the long-burst population, the fraction of binaries that remain bound should be ${\cal R}_{\text{short}}/{\cal R}_{\text{long}}\approx 2$\%--$8$\%. Thus, the SN explosion would disrupt the $92\%$--$98\%$ of NS-NS binaries from BdHNe~II and III. However, the latter dominates the percentage of unbound systems given their much wider pre-SN orbits \cite{2024arXiv240115702B}. Interestingly, this inferred $\sim 1\%$ fraction of survived NS-NS binaries only based on the GRB rates and the BdHN prediction that short GRBs are long GRB descendants agrees with estimates from population synthesis simulations \cite[see, e.g.,][and references therein]{2012ApJ...759...52D, 2014LRR....17....3P, 2015ApJ...806..263D, 2015ApJ...812...24F, 2016ApJ...819..108B}. See also \citet{2019MNRAS.485.5394K, 2022MNRAS.513.3550C, 2022RNAAS...6...13L, 2023MNRAS.522.2029C}, for more recent analyses. All the above has triggered new observational campaigns searching for bound or ejected companions of SN explosions \cite[see, e.g.,][and references therein]{2021MNRAS.505.2485O,2022ApJ...929L..15F,2023ApJ...956L..31M,2023ApJ...949..121C, 2023MNRAS.522.2029C, 2024Natur.625..253C}.

\subsection{GRB redshift distribution}\label{sec:2.2}

\begin{figure}
    \centering
    \includegraphics[width=\hsize,clip]{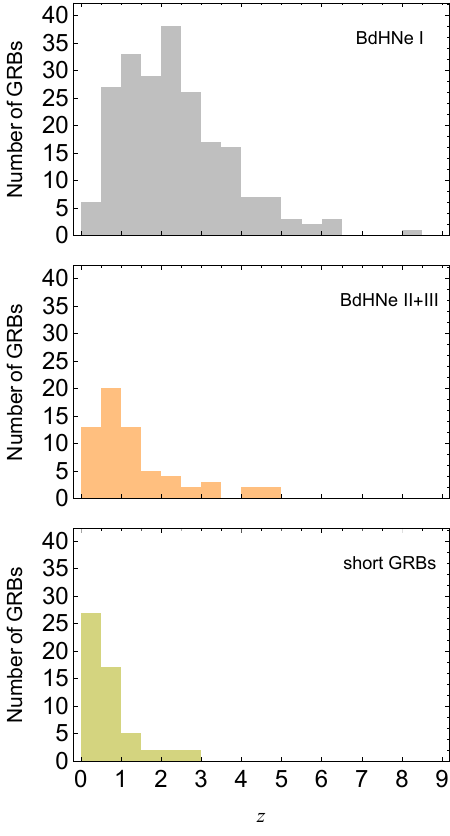}
    \caption{{Distributions of a sample of $335$ GRBs as a function of the cosmological redshift. The sample is divided into three subsamples: BdHNe~I (upper panel, $216$ sources, gray), BdHNe~II+III (middle panel, $64$ sources, orange), and short GRBs (lower panel, $55$ sources, green). This GRB sample is an updated version, with $34$ additional short GRBs until the end of $2023$, of the one considered by \citet{2024ApJ...966..219B}. We refer to Sec.~6 of \citet{2024ApJ...966..219B} for additional details on the definition of the sample.}}
    \label{fig:distributions}
\end{figure}

In \citet{2024ApJ...966..219B}, the redshift distribution of a sample of $301$ GRBs observed by Swift before December 2018 was analyzed. Based on the definition of long GRB types within the BdHN scenario and that of short GRBs, the above Swift sample was subdivided into three subsamples: $216$ BdHNe~I, $64$ BdHNe~II and III, and $21$ short GRBs. The redshift distribution of the BdHNe~I subsample shows a single peak between $z\sim 2$ and $z\sim 2.5$ and a sort of plateau for $0.5 \lesssim z \lesssim 2$. The distribution of the subsample formed by BdHNe~II and III shows a single peak around $z\sim 1$. Therefore, the distribution of BdHN I+II+III has a double-peak structure \cite{2024ApJ...966..219B}, which, as expected, agrees with previous analysis of the long GRB population (see, e.g., \cite{2010MNRAS.406.1944W, 2016ApJ...829....7L}, and Fig. 8 in \citet{2012MNRAS.423.3049G}). The sample of short GRBs shows a single peak at $z\lesssim 0.5$. {In this paper, we updated this GRB sample by considering $34$ additional short GRBs until the end of $2023$. The total number of short GRBs in this new sample is, therefore, $55$, and the total number of GRBs in the entire sample is $335$. Figure \ref{fig:distributions} shows the distributions of the BdHNe~I (upper panel), BdHNe~II+III (middle panel), and short GRBs (lower panel) subsamples.}
It shows the following qualitative features:
\begin{itemize}
    \item The BdHN I population is responsible for the long GRB peak at $z^{I}_p\sim 2$--$2.5$ \cite{2024ApJ...966..219B}. {The BdHN II+III distribution peaks at $z^{\rm II+III}_p\approx 0.72$. One of the reasons for $z^{\rm I}_p > z^{\rm II+III}_p$ is the BdHN I higher energetics, which allows their detection at larger redshifts.}
    \item The distributions of BdHNe~II+III and short GRBs show a similar shape \cite{2024ApJ...966..219B}. The former is wider than the latter, and their peaks occur at slightly different redshifts. {The peak of the short GRB distribution occurs at $z^{\rm short}_p\approx 0.42$, which is lower than $z^{\rm II+III}_p \approx 0.72$ by $\Delta z \approx 0.3$.}
\end{itemize}

We have performed a Kolmogorov-Smirnov test on the relation hypothesis between the BdHN~I, BdHN~II+III and short GRB distributions. The following conclusions can be drawn:
\begin{itemize}
    \item The $p$-value testing the BdHN I and short GRB distribution similarity is $4.5\times 10^{-10}$. This very low value suggests their relationship is unlikely.
    \item The $p$-value testing the BdHN II+III and short GRB distribution similarity is $0.011$. This much larger value indicates similarity. The difference in the position of the peaks dominates the difference in the distributions. In fact, by shifting any of the distributions by the difference of their peaks, $\Delta z\approx0.3$, the $p$-value increases to $\approx 0.35$.
\end{itemize}

The above results agree with our previous conclusions based on the GRB density rates: the observed population of short GRBs appears dominated by NS-NS mergers and not by NS-BH mergers, so it is not evolutionarily connected with the BdHN~I population but with that of BdHNe~II and III, i.e., the latter may form the NS-NS binaries that become the short GRB progenitors. This conclusion finds further support from the estimated merger times. The most recent numerical simulations of the BdHN scenario \cite{2024arXiv240115702B} lead to a wide range of merger timescales $\sim 10^4$--$10^9$ yr (see Fig.~\ref{fig:tauGW} below). The rapidly merging binaries are those of short orbital periods, so they are mostly NS-BH, which have merger times $\tau_{\rm merger}\sim 10$ kyr \cite{2015PhRvL.115w1102F}. As we discussed above, those NS-BH are post-BdHN I products. Thus, given the peak of the BdHN I distribution at $z\sim 2$ and the NS-BH short merging times, these binaries should not be expected to contribute to the short GRB population at low redshifts.  

\subsection{GRB host galaxies and projected offsets}\label{sec:2.3}

Concerning the short GRB host galaxies, \citet{2022ApJ...940...57N} shows that $84\%$ are star-forming, like long GRB hosts. This fraction decreases significantly at low redshift ($z\lesssim 0.25$), in line with galaxy evolution. Interestingly, high-mass galaxies are less abundant among the short GRB hosts than field galaxies, which becomes more evident at $z\gtrsim 0.5$ and more similar to the analogous distribution for long GRB hosts. Moreover, they found evidence for both a short delay-time population, mostly for star-forming hosts at $z>1$, and a long delay-time one, which becomes prevalent at lower redshift in quiescent hosts.

The projected physical offsets from the host galaxy center of short GRBs are, on average, larger than those of long GRBs. Recent work by \citet{2022ApJ...940...56F} including $90$ short GRB host galaxies, the majority of which are robust associations, finds offsets ranging from a fraction of kpc to $\approx 60$ kpc, with a median offset value $5$--$8$~kpc (see also \citealt{2022MNRAS.515.4890O}). These values must be compared with the median value of long GRBs of $1.28$~kpc. Indeed, $90\%$ of long GRB offsets are $<5$~kpc \cite{2016ApJ...817..144B}. 

The above observational properties evidence that, for long and short GRBs to share a common progenitor, the delay time distribution of the compact-object binary mergers must include short and long values. We shall discuss these points in the next section.

\section{Post-BdHN NS-NS/NS-BH time and distance traveled to merger}\label{sec:3}

We have recently presented in \citet{2024arXiv240115702B} a new set of numerical simulations performed with the \textsc{SN-SPH} code  \cite{2006ApJ...643..292F} of the evolution of the binary system from the CO star SN explosion. The code follows the structure evolution of the $\nu$NS and the NS companion as they move and accrete matter from the SN ejecta. The initial setup has been described in detail in \citet{2019ApJ...871...14B} (see also \cite{2022PhRvD.106h3002B}). 

The code tracks the SN ejecta and point-mass particles' position and velocity. The total energy of the evolving $\nu$NS-NS system, $E_{\rm tot}$, is given by the sum of the total kinetic energy relative to the binary's center of mass and the gravitational binding energy. The system is bound if $E_{\rm tot}<0$. In that case, the orbital separation can be determined from the binary total energy, the orbital period from Kepler's law, and the eccentric from the orbital angular momentum \cite[see][for further details]{2024arXiv240115702B}.

To examine the conditions under which the binary remains bound, we perform simulations for various initial orbital periods, keeping fixed the initial mass of the NS companion, $M_{\rm NS,i}=2 M_\odot$, the ZAMS progenitor of the CO star ($M_{\rm zams}=25M_\odot$), and the SN explosion energy. The pre-SN CO star has a total mass of $M_{\rm CO}=6.8~M_\odot$ and leaves a $\nu$NS of $M_{\rm \nu NS,i}=1.8~M_\odot$. Thus, it ejects $M_{\rm ej}\approx 5~M_\odot$ in the SN explosion. We recall that $M_{\rm CO}=M_{\nu \rm NS, i}+M_{\rm ej}$. We record the final values of the $\nu$NS mass, $M_{\rm \nu NS,f}$, the NS companion mass, $M_{\rm NS,f}$, orbital separation, $a_{\rm orb,f}$, orbital period, $P_{\rm orb,f}$, and eccentricity, $e_f$. Another key quantity is the final binary center of mass velocity, $v_{\rm cm, f}$. We end the simulation when most of the ejecta have left the system, i.e., when the mass gravitationally bound to the stars ($\nu$NS and NS) is gravitationally negligible, e.g., $\lesssim 10^{-3}~M_\odot$.

The final total energy of the systems in the simulations is well-fitted by the following polynomial function:
\begin{equation}\label{eq:max_P}
    E_{\rm tot,f}\approx -\frac{1}{2}\frac{G M_{\rm CO}M_{\rm ns, i}}{a_{\rm orb, i}} (a + b x +c x^2),\quad x\equiv \frac{a_{\rm orb,i} P_{\rm orb,i}}{v_{\rm sn}},
\end{equation}
where $v_{\rm sn}=\sqrt{2 E_{\rm sn}/M_{\rm ej}}$ is an indicative average expansion velocity of the SN ejecta of mass $M_{\rm ej}$. For the present binary, $a=0.294$, the constants $b$ and $c$ depend on the SN explosion energy and are listed in Table 2 of \citet{2024arXiv240115702B}. For example, for $E_{\rm sn} = 6.3 \times 10^{50}$ erg, $b = -3.153$ and $c = 5.219$. The maximum initial period for the system to hold bound is obtained by setting the final total energy to zero. In the present example, the energy becomes zero at $x = 0.115$, which implies $P_{\rm orb,max} \approx 7.15$ min.

The final bound systems will be compact binary systems (NS-NS or NS-BH), which will eventually merge through the emission of gravitational waves. The time to merger is given by \cite[see, e.g.,][]{maggiore2007gravitational}
\begin{align}\label{eq:taumerger}
    \tau_{\rm merger} &= \frac{c^5}{G^3}\frac{5}{256} \frac{a_{\rm orb}^4}{\mu M^2} F(e),\\
    F(e) &= \frac{48}{19}\frac{1}{g(e)^4}\int_0^e \frac{g(e)^4 (1-e^2)^{5/2}}{e (1 + \frac{121}{304}e^2)}de,
\end{align}
where $g(e) = e^{12/19}(1-e^2)^{-1}(1+121 e^2/304)^{870/2299}$, being $M=m_1+m_2$, $\mu = m_1 m_2/M$, and $e$ the orbit total mass, reduced mass, and eccentricity.
\begin{figure*}
    \centering
    \includegraphics[width=0.47\hsize,clip]{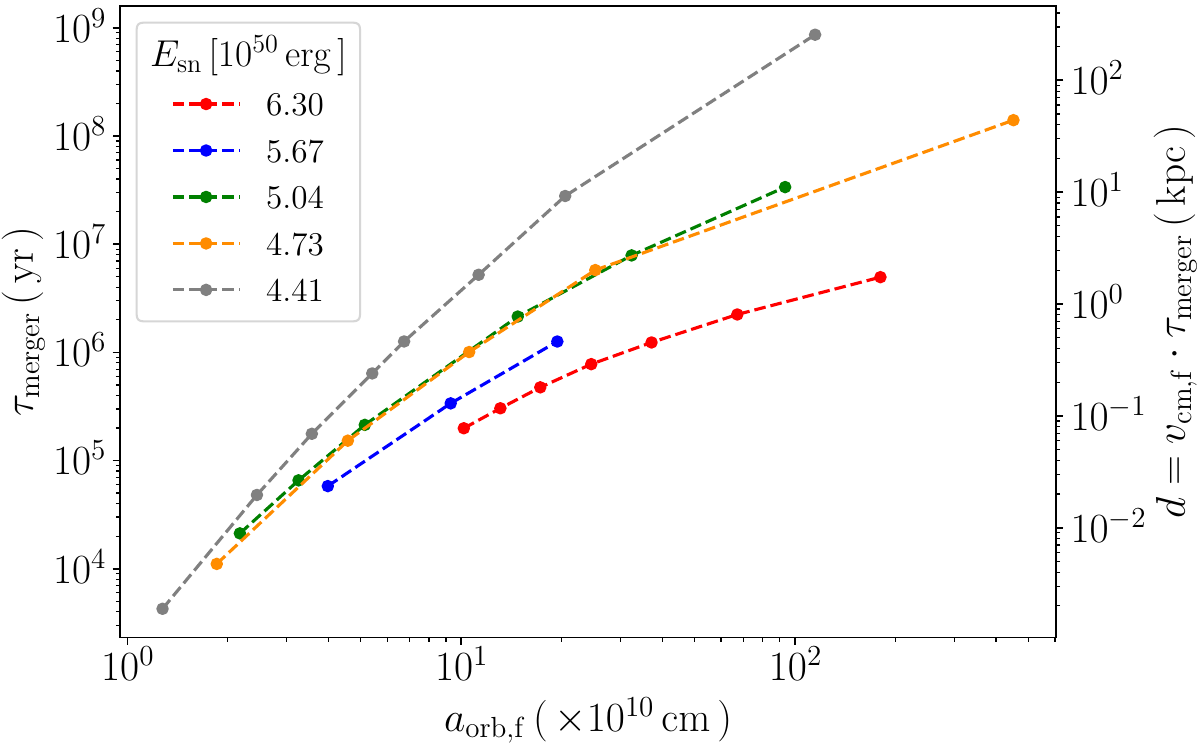}\hspace{1cm}\includegraphics[width=0.47\hsize,clip]{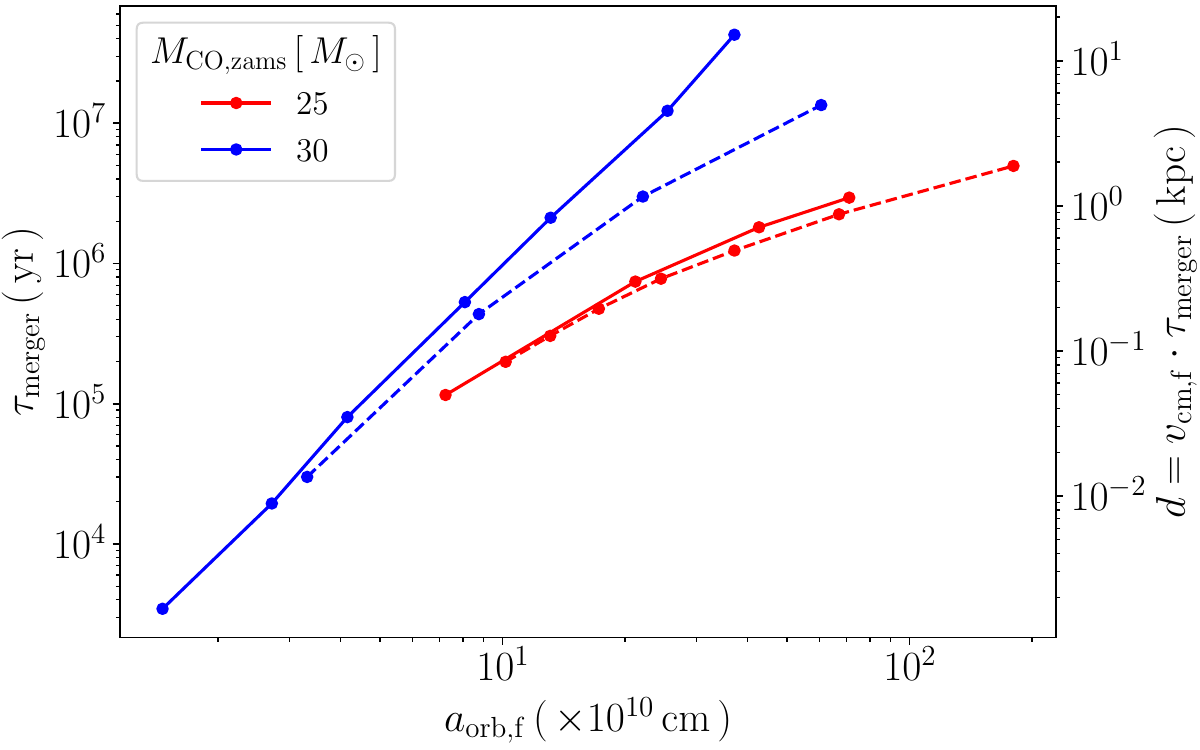}
    \caption{Characteristic merger time by gravitational-wave emission (left axis) and distance travel (right axis) for the binary systems that remain bound (negative total energy) after a BdHN event as a function of the final binary separation. Left: the initial binary comprises a CO-evolved star from a ZAMS progenitor of $M_{\rm zams}=25M_\odot$ and a $2M_\odot$ NS companion and the curves correspond to selected SN explosion energies. Right: Simulations for the SN explosion energy $6.30\times 10^{50}$ erg for two CO-evolved stars from ZAMS progenitors: $M_{\rm zams}=25M_\odot$ (red) and $30M_\odot$ (blue). The dashed (solid) curves correspond to symmetric (asymmetric) SN explosions \cite[see][for details]{2024arXiv240115702B}.}
    \label{fig:tauGW}
\end{figure*}

We have calculated the time to merger from Eq. (\ref{eq:taumerger}), using the parameters obtained from the numerical simulations, i.e., $a_{\rm orb} = a_{\rm orb,f}$, $m_1 = M_{\nu \rm NS,f}$, $m_2 = M_{\rm NS,f}$, and $e = e_f$. With this information, the distance traveled by the newly formed compact object binary from the BdHN event location to the merger site is
\begin{equation}\label{eq:offset}
    d = v_{\rm cm,f}\,\tau_{\rm merger}.
\end{equation}
Figure~\ref{fig:tauGW} shows $\tau_{\rm merger}$ (left axis) and $d$ (right axis) as a function of $a_{\rm orb,f}$. We show the results when the CO star's companion is an NS of $M_{\rm NS, i} = 2M_\odot$, while we adopt two models for the CO star. The first is the model of the previous example, i.e., a CO-evolved star from a ZAMS progenitor of $M_{\rm zams}=25 M_\odot$; $M_{\rm CO} = M_{\nu \rm NS, i} +M_{\rm ej} \approx 6.8 M_\odot$, where $M_{\nu \rm NS, i} \approx 1.8 M_\odot$ and $M_{\rm ej} \approx 5 M_\odot$. The second model is the CO star from a $M_{\rm zams}=30 M_\odot$; $M_{\rm CO} \approx 8.9 M_\odot$, where $M_{\nu \rm NS, i} = 1.7 M_\odot$ and $M_{\rm ej} \approx 7.2 M_\odot$. Each point in each curve corresponds to a different value of the parameter $x$ defined in Eq. (\ref{eq:max_P}), so for fixed initial component masses, ejecta mass, and SN explosion energy, it explores a range of orbital periods $P_{\rm orb, i}$ (or, equivalently, $a_{\rm orb, i}$). In the right panel plot, we compare the results for a symmetric and asymmetric SN explosion of the same energy.  

For the various SN explosion energies, the left panel of Fig.~\ref{fig:tauGW} shows a range of merger times $\tau_{\rm merger} = 10^4$--$10^9$~yr. Correspondingly, we obtain systemic velocities $v_{\rm cm,f}\sim 10$--$100$~km s$^{-1}$ for those newly-formed binaries. From the above, we find that the distance traveled by these binaries (NS-NS or NS-BH) after the BdHN event ranges $d = 0.01$--$100$ kpc.

The measured projected offsets of long and short GRBs in the host galaxies differ about one order of magnitude (see \citealp{2022ApJ...940...56F} and section~\ref{sec:2.2}). While most long GRBs have offsets $<5$ kpc, with a median value $\sim 1$ kpc, short GRBs show an equally broad distribution but shifted to larger values by about one decade, that is, from a fraction of kpc to $\approx 70$ kpc. The short GRB offset median is $\approx 8$ kpc or $\approx 5$ kpc for the golden sample of the most robust associations. The offsets of the short GRBs in the sample of Fig. \ref{fig:distributions} are $0.15$--$70.19$ kpc. This range of values strikingly agrees with that obtained for the distance traveled by the NS-NS and NS-BH binaries produced by BdHNe. 

{It is worth mentioning that the above conclusions have been obtained within the model's hypotheses and are limited to the parameter space we have explored. Such a parameter space (e.g., CO star mass and orbital period) is not arbitrary; it corresponds to the conditions that, from our simulations, lead to the three subclasses of BdHNe (I, II, III). However, these conditions may vary according to the various physical conditions in population synthesis simulations leading to the pre-BdHN CO-NS binaries. Such simulations are still missing in the literature and represent an interesting new research topic.}

\section{Discussion and conclusions}\label{sec:5}

We have reached the following conclusions:

\begin{enumerate}
    \item \textbf{GRB rates}. The inequality ${\cal R}_{\rm short} < {\cal R}_{\rm long}$ is explained as follows (see section \ref{sec:2.1}). First and foremost, the short GRB is dominated by NS-NS mergers, and only a subset of the BdHNe can produce NS-NS (BdHNe II and III). Thus, the subset leading to short GRBs is given by the BdHNe II and III that lead to bound NS-NS binaries \cite{2024arXiv240115702B}. Further, BdHNe I lead to NS-BH binaries. These binaries can produce short GRBs only if the BH is low enough mass; otherwise, tidal disruption of the NS by the BH is more likely to occur.
    \item \textbf{Redshift distribution}. 
    
    {First, we have shown in section \ref{sec:2.2} that $z^{\rm I}_p (\approx 2-2.5) > z^{\rm II+III}_p (\approx 0.72)$ (see also Fig. \ref{fig:distributions}), which reflects the higher energetics of the BdHN I relative to BdHN II and III that allows their observation at higher redshifts. Then, we showed that the short GRB distribution peaks at $z^{\rm short}_p \approx 0.42$. The inequality $z^{\rm short}_p \ll z^{\rm I}_p$ suggests that BdHN I remnant binaries have a negligible role in the distribution of short GRBs. Indeed, in the BdHN scenario, BdHNe I produce compact-orbit NS-BH binaries, rapidly merging on timescales $<10^5$ yr \cite{2015PhRvL.115w1102F}. At the peak redshift of the BdHN I distribution, $z^I_p \approx 2$--$2.5$, such a timescale implies a negligible redshift interval, so their contribution at $z^{\rm short}_p \approx 0.42$ is negligible. On the other hand, the distribution of BdHN II+III shows similarities with that of the short GRBs, and $z^{\rm II+III}_p \approx 0.72$, which differs from $z^{\rm short}_p$ by $\Delta z = 0.3$. The merger timescales of NS-NS products by BdHN II and III (see Fig. \ref{fig:tauGW}) could explain the time delay (redshift difference) between the two distributions. The above analysis suggests a link between the NS-NS remnant binaries from BdHN II and III as possible progenitors of the short GRBs. Thus, further detailed calculations are needed to deepen this connection, such as simulating the merger time-delay distribution accounting for the occurrence rate and intrinsic distribution of binary periods} at different redshifts and the cosmological expansion. Such a calculation goes beyond the exploratory character of the present article and is left for future analyses.
    \item \textbf{Host galaxies}. Short-GRB host stellar-population ages support the picture of a short delay-time population within young and star-forming galaxies at $z>0.25$, along with a long delay-time population which characterizes older and quiescent galaxies at lower $z$ \cite{2022ApJ...940...57N}. The above observations suggest compact-orbit NS-NS binaries should be more abundant in the former galaxies, while wide-orbit NS-NS binaries dominate in the latter. This suggestive information deserves further attention from combined cosmology and population synthesis models, which, combined with the BdHN simulations, could be used to estimate the expected galactocentric offsets and circum-merger conditions for NS-NS merging systems (see, e.g., \cite{2018ApJ...865...27W}).
    \item \textbf{Galactocentric offsets}: The NS-NS produced by BdHNe II and III have a distribution of binary periods, eccentricities, and systemic velocities, which predict a wide distribution of {systemic velocities $10$--$100$ km s$^{-1}$ and merger times $10^4$--$10^9$ yr, leading to} distances of $0.01$-$100$ kpc traveled by these systems {from the BdHN site} to their merger site at which the short GRBs are expected to be produced (see Fig. \ref{fig:tauGW}). In the BdHN scenario, this distance traveled by the post-BdHN binary directly measures the distance separating the long and short GRB occurrence sites. Therefore, our modeling does not give information on the offset of the long or the short GRB but on their relative offset. Indeed, most long GRBs have offsets $<5$ kpc, while short GRB offsets span from a fraction of kpc to $\approx 70$ kpc. This difference in the offset of about a decade agrees with the BdHN numerical simulations presented here. 
\end{enumerate}

There are additional consequences of the present scenario. Current distributions of merger times and large systemic post-formation velocities are in tension with observations of short GRBs in dwarf galaxies. The velocities larger than the galaxy escape velocities and the long merger times predict offsets larger than observed would impede the r-process enrichment of the galaxy~\cite{2024ApJ...962....5N}. In this regard, our results imply two possibilities. First, a population of short-merger-time binaries ($< 100$~kyr) do not have time to move outside the dwarf galaxy, even for velocities larger than the galaxy's escape velocity. Second, there are binaries with longer merger times but with velocities lower than the galaxy's escape velocity. The present results, combined with future detailed population studies, may determine the relative relevance of these systems to explain these observations.

In summary, we have shown that observations of the GRB density rates and density distribution, the host galaxy types, and the sources' projected position offsets agree with the expectations from the BdHN scenario and numerical simulations. This constitutes a strong test of the surprising conclusion, as it may sound: short GRBs are long GRB descendants.

All the above implies, at the same time, the binary progenitor nature of long GRBs and, consequently, the associated preceding binary stellar evolution. Therefore, further theoretical and observational scrutiny from the GRB, X-ray binaries, population synthesis, stellar evolution, and cosmology communities is highly encouraged.

\acknowledgments{
The simulations were performed on LANL HPC resources provided under the Institutional Computing program.}


\end{document}